\documentstyle[12pt]{article}
\begin{document}

 \title{A note on causality in the bulk \\ and stability on the boundary}
 
\author{ 
Jan Troost 
\thanks{troost@mit.edu; Preprint MIT-CTP-3407}
\\      Center for Theoretical Physics \\  MIT    77 Mass Ave \\ Cambridge, MA 02139 USA  
   }

\maketitle

 \abstract{
By carefully analyzing the radial part of the wave-equation for a scalar field in $AdS_{d+1}$,
we show that for a particular range of boundary conditions on the scalar field, the radial 
spectrum contains a bound state. Using the AdS/CFT correspondence, we interpret this
peculiar phenomenon as being dual to an unstable double trace deformation
of the boundary conformal field theory. We thus show how the bulk theory 
holographically detects whether a boundary perturbation is stable.}

\section{Introduction}
In the study of the $AdS/CFT$ correspondence \cite{Maldacena:1997re}\cite{Gubser:1998bc}\cite{Witten:1998qj}, 
we have learned to translate
holographically many characteristics of the $AdS_{d+1}$ bulk gravitational theory  into features
of the boundary conformal field theory, and vice versa. It is our intention in this paper
to add an entry to the holographic dictionary. 
It was shown previously that a perturbation of the boundary conformal field theory with
a double trace operator (see e.g. \cite{Aharony:2001pa}), can be mimicked in the bulk $AdS_{d+1}$ theory 
by imposing particular
radial boundary conditions on the dual scalar field \cite{Witten:2001ua}\cite{Berkooz:2002ug}
(see also \cite{Klebanov:1999tb}\cite{Minces:2001zy}\cite{Muck:2002gm}\cite{Minces:2002wp}\cite{Sever:2002fk}). 
Depending on the sign of the coefficient of the perturbation,
the boundary theory will be stable or unstable after perturbation \cite{Witten:2001ua}. 
We ask  how the bulk theory detects
an unstable boundary deformation.
Technically, the answer to this innocuous question lies hidden in a careful determination
of the spectrum of the radial part of the wave equation in $AdS_{d+1}$, to which we turn.
\section{Set-up}
We write the $AdS_{d+1}$ metric
in Poincare coordinates ($u,x^{\mu})$ as:
\begin{eqnarray}
ds^2 &=& \frac{L^2}{u^2} (du^2+(d x^{\mu})^2)
\end{eqnarray}
where $x^{\mu}$ parametrizes a $d$-dimensional Minkowski space
(with mostly plus signature)
and the $u$-coordinate runs over the interval $]0,\infty [$.
The scale $L$ sets the radius of the $AdS_{d+1}$ space (and
the negative cosmological constant). In the following we set $L=1$.
The $u$-coordinate is intuitively thought of as the inverse of
a radial coordinate ($u=\frac{1}{r}$) with the boundary of the 
$AdS$ space at $u=0$.
The wave-equation for a massive minimally coupled scalar field $\Phi$ is:
\begin{eqnarray}
(\Box-m^2) \Phi(u,x^{\mu}) &=& 0.
\end{eqnarray}
where $\Box$ denotes the wave-operator in $AdS_{d+1}$.
We will analyze the solutions to the wave-equation by using the method
of separation of variables. We expand the solutions in 
multiples of Minkowski plane waves $\phi( x^{\mu} )= e^{ i k_{\mu} x^{\mu} }$
as $\Phi=e^{ i k_{\mu} x^{\mu} } 
u^{\frac{d-1}{2}} f(u)$.
The Minkowski and the radial part of the wave-function then satisfy
the equations:
\begin{eqnarray}
(\Box_M-\lambda) \phi(x^{\mu}) &=&0 \nonumber \\
-f''(u) + \frac{\nu^2-1/4}{u^2} f(u) &=& \lambda f(u),
\label{waves}
\end{eqnarray}
where $\Box_M$ denotes the wave-operator in Minkowski space,
and $\lambda=-k^{\mu} k_{\mu}$ can be thought off as a Minkowksi mass
squared. The parameter $\nu$ is defined by the formula $\nu=\sqrt{m^2+d^2/4}$.
The radial part of the wave-equation corresponds to an interesting,
non-trivial Sturm-Liouville problem which we need to analyze in detail.
\section{Sturm-Liouville}
The key to the bulk interpretation of an unstable double trace boundary
deformation will be the spectrum of the radial Sturm-Liouville 
problem. In this section, we will determine the spectrum rigorously.

The potential for the radial problem is $q(u)=\frac{\nu^2-1/4}{u^2}$.
It is well-known that at the end of the half-line (at $u=0$), we need
to specify boundary conditions for the Sturm-Liouville problem when
$0 \le \nu <1$, because then the potential is not steep enough to
provide for a single dynamical solution to the (one-dimensional)
scattering problem.
For these values of the mass squared \cite{Breitenlohner:bm}\cite{Breitenlohner:jf},
 we need to specify how a wave reflects
from the end point of the half-line (i.e. at radial infinity in $AdS_{d+1}$)
 to determine its full evolution.
(I.e. we need to choose a self-adjoint extension of the radial differential
operator \cite{RS}. See also \cite{Bertola:1999vq}\cite{Satoh:2002bc}.) 
We restrict to the interesting mass squared range
$-d^2/4 \le m^2 < 1-d^2/4$ where $0 \le \nu <1$. Note that we take the
scalar field to satisfy the Breitenlohner-Freedman lower 
bound \cite{Breitenlohner:bm}\cite{Breitenlohner:jf}.

For the analysis of the Sturm-Liouville problem we follow the rigorous textbook treatment in
\cite{T}. We
refer to \cite{T} for a pedagogical explanation of the nomenclature and
techniques involved in the following subsections.
The two ends of the half-line
are singular points for the Sturm-Liouville problem. To treat them
rigorously, we 
cut the half-line into an interval $ ] 0,\epsilon [ $ 
and another half-line $ ] \epsilon , \infty [$ (see e.g. \cite{T} section
2.18). Physically, we introduce an 
infrared bulk cut-off $\epsilon$ and study the behavior at radial infinity
(small $u$)
separately from the behavior of the wave-function in the far interior (large $u$). 
At the end, we will combine the solutions to the separate Sturm-Liouville problems.
The following subsections contain the necessary
technical manipulations to obtain the spectrum, which is summarized at the end
of subsection \ref{full}.
\subsection{Near the boundary}
On the interval $ ] 0,\epsilon [ $, there are two solutions to the
radial equation which are both
quadratically integrable. We are in the limit circle case for the
singularity at $0$ (see e.g. \cite{T} p.23-25).
Given two normalizable solutions $\phi$ and $\theta$, we then
have to specify a boundary condition to pick a unique solution $\psi_1=
\theta+m_1 (\lambda) \phi$ that satisfies these boundary conditions.
Thus, the function $m_1(\lambda)$, which encodes the spectrum, depends
on the boundary conditions.
Some technical details follow.
We define $\phi$ and $\theta$ to be the solutions to the differential equation satisfying the
boundary conditions:
\begin{eqnarray}
\phi(\epsilon, \lambda) =0  &{}&  \phi'(\epsilon,\lambda) =-1 \nonumber \\
\theta(\epsilon,\lambda) =1 &{}& \theta'(\epsilon,\lambda) =0.
\end{eqnarray}
They are given by
(with $\lambda = s^2$):
\begin{eqnarray}
\phi(u,\lambda) &=& \frac{\pi}{2} u^{1/2} \epsilon^{1/2} (J_{\nu} (us) 
Y_{\nu}(\epsilon s)-Y_{\nu}(us) J_{\nu}(\epsilon s)) \nonumber \\
\theta(u,\lambda) &=& \frac{\pi}{2} u^{1/2} \epsilon^{1/2} s (J_{\nu} (us) 
Y_{\nu}'(\epsilon s)-Y_{\nu}(us) J_{\nu}'(\epsilon s)) + \frac{\phi(u,\lambda)}{2 \epsilon}
\label{thetaphi}
\end{eqnarray}
as is easily verified using the fact that $J_{\nu}(x) Y_{\nu}'(x)-J_{\nu}'(x) Y_{\nu}(x)=2/(\pi x)$.
 We suppose in the following
that $\nu \neq 1/2$ and
$0<\nu<1$. (The case $\nu=0$ deserves a separate treatment. See e.g. \cite{Witten:2001ua}\cite{T}.)
We consider a solution with boundary condition $\mbox{cot} \delta = -\frac{f'(\epsilon)}{f(\epsilon)}$
 and define the limit circle as the limit of the
circles (circumscribed by varying $\mbox{cot} \delta$) as $\epsilon \rightarrow 0$:
\begin{eqnarray}
l(\epsilon)&=& -\frac{\theta(\epsilon,\lambda) \mbox{cot} \delta + \theta'(\epsilon,\lambda)}{
\phi(\epsilon,\lambda) \mbox{cot} \delta  + \phi'(\epsilon,\lambda)}.
\end{eqnarray}
When we take the limit $\epsilon \rightarrow 0$, with $\lambda=s^2$ fixed,
we find it useful to define the quantity $c$ by the formula:
\begin{eqnarray}
\frac{\epsilon^{1/2-\nu} \mbox{cot} \delta+ (1/2-\nu) \epsilon^{-1/2-\nu}}{2^{-\nu}
\Gamma(1-\nu)} & \equiv & c \frac{\epsilon^{1/2+\nu} \mbox{cot} \delta+
 (1/2+\nu) \epsilon^{-1/2+\nu}}{2^{\nu}
\Gamma(1+\nu)}.
\end{eqnarray}
Using the fact that $\mbox{cot} \delta = O(\frac{1}{\epsilon})$ in this limit, we find that
(for the specified range of $\nu$) $l(\lambda)$ asymptotes to $m_1(\lambda)$:
\begin{eqnarray}
m_1(\lambda) &=& -s
 \frac{c s^{-\nu} J_{\nu}'(\epsilon s)-
s^{\nu} J_{-\nu}'(\epsilon s)}{c s^{-\nu} J_{\nu}(\epsilon s)-s^{\nu} J_{-\nu}(\epsilon s)}- \frac{1}{2\epsilon}.
\end{eqnarray}
Note that $c$ parametrizes the boundary condition at radial infinity, via $\mbox{cot} \delta$ and that
$m_1(\lambda)$ encodes the chosen normalizable solution as well as the spectrum of the Sturm-Liouville
operator with the given boundary conditions (see e.g. \cite{T} section 3.9).
\subsection{The interior}
We now turn to the analysis of the Sturm-Liouville problem on the half-line $]\epsilon, \infty[$.
When we regularize the Sturm-Liouville problem
by adding a small positive imaginary part to $\lambda$, the one normalizable solution is proportional 
to the first Hankel  function $H^{(1)}_{\nu}(u s)$. (Physically, the regularisation procedure
picks out a particular contour prescription for the propagator \cite{Satoh:2002bc}.)
That leads to the expression for the function $m_2(\lambda)$ (as can easily be checked using
the fact that the normalizable solution is $\psi_2=\theta + m_2 \phi$, and formulas (\ref{thetaphi})):
\begin{eqnarray}
m_2(\lambda) &=& -s \frac{{H_{\nu}^{(1)}}'(\epsilon s)}{H_{\nu}^{(1)}(\epsilon s)} - \frac{1}{2 \epsilon}.
\end{eqnarray}

\subsection{The full problem}
\label{full}
Next we combine our results to treat the
full Sturm-Liouville problem on the half-infinite interval with the two singular ends. 
We take $\epsilon$ as the 
basic point. We have found the wave functions
\begin{eqnarray}
\psi_1 &=& \frac{u^{1/2}}{\epsilon^{1/2}} 
\frac{cJ_{\nu}(us)-s^{2 \nu} J_{\nu}(us)}{cJ_{\nu}(\epsilon s)-s^{2 \nu}J_{-\nu}(\epsilon s)}
\nonumber \\
\psi_2 &=& \frac{u^{1/2} H^{(1)}(u s)}{\epsilon^{1/2} H^{(1)}(\epsilon s)}.
\end{eqnarray}
To obtain the spectrum\footnote{For the general theory see \cite{T} p. 52-53 and section
3.9.}, we need to study the function $m_1(\lambda)-m_2(\lambda)$.
For $\lambda>0$ ($s$ real and positive) we obtain:
\begin{eqnarray}
- Im \frac{1}{m_1(\lambda)-m_2(\lambda)} &=& \frac{\pi \epsilon}{2}
\frac{(c J_{\nu}(\epsilon s)-s^{2 \nu} J_{-\nu}(\epsilon s))^2}{c^2-2 c s^{2 \nu} \cos \nu \pi+s^{4 \nu}}
\end{eqnarray}
such that we find a continuous spectrum (since the above expression varies continuously
with $\lambda$ and does not contain poles) for $\lambda>0$.
For $\lambda<0$ ($\lambda=s^2$, $s=it$ and $t$ real and positive) we find that:
\begin{eqnarray}
-\frac{1}{m_1(\lambda)-m_2(\lambda)} &=& \frac{\epsilon K_{\nu}(\epsilon t)
(c I_{\nu}(\epsilon t)-t^{2 \nu} I_{-\nu}(\epsilon t))}{c-t^{2 \nu}}
\end{eqnarray}
is real. Now, if $c<0$, the last expression is continuous with no poles and gives
no contribution to the spectrum.
When $c>0$, we find a pole, which indicates another (discrete)
eigenvalue in the spectrum. The new eigenvalue $\lambda=-c^{1/ \nu}$ 
is central to our paper.

As a check on our formalism, note that the 
Green functions $G(x,y;\lambda)$ can be written down as a combination
of $m_{1,2}$ and $\psi_{1,2}$:
\begin{eqnarray}
G(x,y; \lambda) &=& \frac{\psi_2(x,\lambda) \psi_1(y,\lambda)}{m_2(\lambda)-m_1(\lambda)} \qquad
\mbox{for} \quad (y \le x) \nonumber \\
                &=& \frac{\pi i }{2} \frac{1}{c-s^{2 \nu} e^{- i \nu \pi}}
x^{1/2} H^{(1)}_{\nu}(xs) y^{1/2} (c J_{\nu}(ys)-s^{2 \nu} J_{-\nu}(ys)). 
\label{green}
\end{eqnarray}
It is understood in the previous expression that the role of $\psi_1$ and $\psi_2$ is
reversed for $y > x$.
We note that the Green function agrees with the radial part of the Green function 
computed in \cite{Gubser:2002zh} (equation (32)),
after using the formulas $\pi i H_{\nu}^{(1)}(iz) = 2 e^{-i \nu \pi /2} K_{\nu}(z)$
and $I_{\nu}(z)= e^{-i \nu \pi / 2 } J_{\nu}(iz)$ which connect different forms of the
(modified) Bessel functions.

However, to obtain the Green function for the full bulk problem in $AdS_{d+1}$, it is crucial to know
the precise spectrum of eigenvalues $\lambda$ over which we have to integrate the product of the
Minkowski and radial Green functions. That is why we embarked on the long derivation of the 
spectrum above. We have found a continuous spectrum, with an extra discrete eigenvalue
when $c>0$, i.e. an extra discrete eigenvalue appears for a particular range of boundary conditions.
The bound state is directly associated to the appearance of a pole
in the Green function (\ref{green}).
\section{Holographic translation}
We demonstrate now that the peculiar feature of the bulk theory uncovered above, provides
the answer to the question we posed in the introduction.
Recall that in the mass range $-d^2/4 < m^2 < 1-d^2/4$, we can approximate the scalar field near
the boundary by:
\begin{eqnarray}
\Phi(u) & \approx & \alpha u^{\frac{d}{2}-\nu} + \beta u^{\frac{d}{2}+\nu}. 
\end{eqnarray} 
We show how the values for $c$ translate into properties of the coefficients
$\alpha$ and $\beta$. We define the following quantity which encodes the boundary
condition on the scalar field, but does not scale with
$\epsilon$:
\begin{eqnarray}
A &=& \epsilon \, \mbox{cot} \delta .
\end{eqnarray}
It satisfies the relation:
\begin{eqnarray}
c = \epsilon^{-2 \nu} 2^{2 \nu} \frac{\Gamma(1+\nu)}{\Gamma(1- \nu)} \frac{1/2-\nu+A}{1/2+\nu+A}.
\end{eqnarray}
Since we have that $\epsilon>0$ and $0 < \nu <1$, it is easy to check that
$c$ is positive when $A > \nu -1/2$ and when $A < - \nu -1/2$, and that $c$ is
negative for $ - \nu -1/2 < A < \nu -1/2 $. 

Recall that $A=- \epsilon \frac{f'(\epsilon)}{f(\epsilon)}$ (and
$f(u)= \alpha u^{1/2-\nu}+\beta u^{1/2+\nu}$), such that
we find that for the boundary conditions consistent with the full $AdS_{d+1}$
isometry group (i.e. boundary conformal 
symmetry), i.e. $\beta=0$ or $\alpha=0$, we obtain the limiting values
$A=\nu-1/2$ or $A=-\nu-1/2$ respectively. When $\alpha \neq 0$, we can show
that $A> \nu-1/2$ for small $\epsilon$ and $\beta/\alpha>0$ while 
$A<\nu-1/2$ for $\beta/\alpha<0$. In general we find that there is no bound
state for $\beta/\alpha>0$ and $c<0$, while there is a bound state for $c>0$ or in other
words $\beta/\alpha<0$. (Note that for $\alpha \neq 0$ we find that 
$c= -2^{2\nu} \frac{\Gamma(1+\nu)}{\Gamma(1-\nu)} \frac{\beta}{\alpha}$
for small $\epsilon$.)

We note in passing that is not difficult to formulate an action principle (with a
boundary term) that gives rise to the boundary conditions we analyzed 
(see e.g. \cite{Sever:2002fk}). 

\subsection*{Interpretation}
In \cite{Witten:2001ua} it was argued that the boundary condition $\alpha=g \beta$ for the
quantization of a scalar field in $AdS$ is dual to adding a double
trace deformation for the operator coupling to the field $\Phi$
in the boundary conformal field theory. The sign of
$g$ determines whether the boundary perturbation is stable or not. 
We have now
seen that the sign appearing in the boundary condition also determines whether
the bulk wave equation allows for a bound state with negative energy in the
corresponding radial Schr\"odinger problem. The negative energy bound state
corresponds to a wave in the Minkowski slices of the Poincare patch with a
disturbing negative (Minkowski) mass squared $\lambda$ (see (\ref{waves})). This is the
bulk holographic signal for the instability of the boundary theory.
\section{Conclusions}
We added an entry to the $AdS/CFT$ dictionary. 
An instability in the boundary conformal field theory,
caused by perturbing it with a double trace deformation with 
the ``wrong'' sign (see e.g. appendix of \cite{Gross:jv}), is
detected in the bulk theory by the appearance of a solution to the wave equation
with tachyonic behavior in a Minkowski slice. Note that the tachyonic behavior
appeared due to the boundary conditions solely and that the Breitenlohner-Freedman is
satisfied by the mass squared of the field. This is a perhaps surprising
feature of the $AdS_{d+1}$ scalar wave equation (and thus of the quantization
of scalar fields in $AdS_{d+1}$.) The
careful analysis of the Sturm-Liouville differential equation for the radial part
of the scalar field was instrumental in laying bare this quaint
feature of the $AdS/CFT$ correspondence. 

The quantization of a scalar field in $AdS_{d+1}$ with non-conformal, stable boundary
conditions found applications in the analysis of
the flow of a bulk analogue of the central charge in the boundary conformal field 
theory \cite{Gubser:2002zh} and the Legendre transform between conformal fixed points
\cite{Gubser:2002vv}. (See also \cite{Nojiri:2003zq}.) 
We have shown that the stable quantization used in that analysis seems to be the
only intuitively acceptable one from a bulk perspective
(although more general boundary conditions exist). 

A bulk theory is specified not
only by the background geometry, but also by the boundary conditions on the propagating
fields. Beyond analyzing which bulk geometries are allowed (see e.g.\cite{Gubser:2000nd}),
one has to specify which boundary conditions are acceptable. One can read our analysis
as specifying the set of allowed boundary conditions in the simple case of pure $AdS_{d+1}$.

It is illuminating 
to see how the instability of the boundary theory is translated in acausal properties
of the bulk theory. In general, it would be interesting to understand still better
the role of causality in the $AdS/CFT$ dictionary,
which is intimately related to getting a grip on truely Lorentzian aspects of the 
$AdS/CFT$ correspondence (see e.g.\cite{Horowitz:1999gf} and
references thereto). The basic mechanism that we identified is an element in that 
broader study.

\section*{Acknowledgments}
Thanks to all the members of the CTP and especially
to Ami Hanany and Pavlos Kazakopoulos for useful discussions. 
 Our
research was supported by the U.S. Department of Energy under cooperative 
research agreement \# DE-FC02-94ER40818.

\end{document}